\documentclass[12pt]{article}   %
\usepackage{psfig}         %
     \newcommand{\ds}{\displaystyle}
     
\begin{document}\hbadness=10000
\markboth{J. Rafelski, J. Letessier and A. Tounsi}{HADRONIC SIGNATURES OF DECONFINEMENT}
\title{HADRONIC SIGNATURES OF DECONFINEMENT \\
IN RELATIVISTIC NUCLEAR COLLISIONS\footnotemark
\footnotetext{Presented by Jan Rafelski 
at XXXVII Cracow School of Theoretical Physics,\\
Zakopane May 30 -- June 10, 1997}
}
\author{\large\bf
Jan Rafelski\thanks{\em Support by 
US-Department of Energy 
under grant DE-FG03-95ER40937.
}, \,Jean Letessier\  and Ahmed Tounsi
\\ $\ $\\
 Department of Physics, University of Arizona, Tucson, AZ 85721\\$\ $\\ Laboratoire de Physique Th\'eorique et Hautes Energies\thanks{\em
Unit\'e  associ\'ee au CNRS UA 280.}\\
Universit\'e Paris 7, 2 place Jussieu, F--75251 Cedex 05.\\
}
\date{September 30, 1997}
\maketitle
\vskip -10.2cm 
\vskip -0.5cm \noindent 
{Submitted to {\it Acta Physica Polonica B
}\hfill PAR/LPTHE/97-43}\\
\vskip 10.2cm
\centerline{Dedicated to {Professor Wieslaw Czyz} 
on occasion of his 70th birthday}
\begin{abstract}
We describe the  remarkable accomplishments of
the current heavy ion Pb--Pb collision experiments 
involving strange 
particle production, carried out at 158A GeV at CERN--SPS. 
These experimental results, together with the earlier
200A GeV S-induced reactions, imply that, at central  rapidity, 
a novel mechanism of strangeness production arises, accompanied by 
excess entropy formation. We argue that:\\
$\bullet$ these  results are consistent 
with the formation of a space-time localized, 
highly excited, dense state of matter;\\ 
$\bullet$ the freeze-out properties of strange hadrons 
are suggestive of the  formation of 
a color-deconfined, thermally and  nearly chemically 
equilibrated phase, which provides at present the only
comprehensive framework to describe all experimental data;\\ 
$\bullet$ the  matter fireball is undergoing  a 
transverse expansion  with nearly the velocity of sound
of relativistic matter; longitudinal expansion is not 
in the  scaling regime.\\ 
\indent We present a first analysis of the recent Pb--Pb results
and discuss several alternative reaction scenarios.
We evaluate quantitatively strangeness production  in the 
deconfined quark-gluon phase and obtain yields 
 in agreement with the experimental
observations made in 200A GeV S--W  and 158A GeV Pb--Pb interactions. 
We also present a qualitative discussion 
of $J/\!\Psi$ results consistent with our understanding of 
strange particle results.\\ \ \\
PACS numbers: {25.75.+, 12.38.Mh, 24.85.+p}
\end{abstract} 

\section{Hadronic Probes of QGP}
For many years now, we have searched to understand what happens
to nuclear matter when it is perturbed by a high energy 
hadron \cite{czyz1}. Our present day interest originates
in the hope and expectation that  novel physical 
phenomena arise in collisions of heaviest nuclei, at center of
momentum energy exceeding many times the  nuclear rest mass.
Specifically, it 
is believed today that, in such collisions, we compress and
significantly excite the nuclear matter of which the nuclei are
made, ultimately dissolving the quarks confined in hadrons
into a bizarre new liquid \cite{Car73}.

The present day experimental study of dense hadronic matter 
formed in relativistic nuclear  collisions is completely 
dominated by the intense search for our modern
times nuclear Holy Grail, the quark-gluon plasma (QGP). 
QGP state must exist according to simple interpretation of
 quantum chromodynamics (QCD), the theory  
of strong interactions. In short, when temperature 
exceeds about 150--200 MeV, the 
structured  QCD vacuum  melts, individual hadrons dissociate 
and fuse \cite{HR80}  and a  macroscopic space time 
region, filled with free, movable quarks and gluons is formed.

At low temperature, in the confining vacuum phase, 
only individual  hadronic particles such as 
$\pi$, $\rho$, $N$, $\Delta$, {\it etc.}, can exist, they are 
in our cold world the condensation drops of
the deconfined state. The great variety of observed 
hadrons  results from the complex nature of 
the QCD interactions and it indirectly implies that 
the structure and properties
of highly excited hadronic matter could be very diverse.

We will compare the experimental data with reaction
models invoking two most opposite states of matter:\\
$\bullet$ 
the conventional, confined phase we shall call hadronic gas (HG),
made of  hadronic particles of different type in local 
thermal equilibrium, with masses and
degeneracies in most cases well known.
 Along with Hagedorn  \cite{HAG},
we  subsume that particle--particle strong 
interactions in HG are accounted  
for by giving the short lived hadronic resonances 
the status of independent statistical fractions;\\
$\bullet$ 
the `melted' QGP phase 
consisting of a nearly free gas of quarks and
gluons, interacting perturbatively, an approach which is properly
justified only in the limit of ultra high energy densities, as seen, 
{\it e.g.}, in  the framework of finite temperature lattice gauge
theory QCD simulations  \cite{lattice}.

Since in the collision of large  nuclei,  the highly dense state is 
formed for a rather short time of magnitude $2R/c$, where $R$ is the nuclear
radius, one of the major challenges 
has been to identify suitable physical observables of deconfinement. 
This difficult problem of detecting reliably the formation of an 
unknown  phase of matter, existing only $0.5\cdot 10^{-22}$sec, has not been
completely resolved today. The electromagnetic probes involving directly
produced photons and dileptons are witnesses to the earliest moments 
of the reaction, but their production rates are in general very small.
Indeed, the direct photon signal remains up to this day undetected, since
the experimental $\gamma$
yield is dominated by the secondary processes $\pi^0\to \gamma+\gamma$.
The dilepton yield is also mostly resulting from meson decays, but 
there are some kinematic regions where this background yield is very small. 
Moreover, the dilepton spectrum gives interesting insights about the 
vector meson yields and their variation with experimental conditions. We
will briefly address here  this very interesting observable
when considering $J\!/\!\Psi$ production in the final section, concentrating
instead on directly detected  hadronic yields and spectra. In passing
we note that 
aside of spectra of hadronic particle, one can also relatively easily measure
particle correlations. The two particle HBT correlation measurement
is widely used to determine the geometric properties of the central
fireball \cite {Bay97}, and the results agree with a reaction picture
between the nuclei relying on geometric considerations.

In this work, we primarily update our comprehensive presentation 
\cite{acta96} of strangeness (charm) and entropy as signature 
of the deconfined  phase. 
Strangeness and entropy are good
 observables since both will
 be preserved  by `reasonable' evolution scenarios
of the dense matter fireball: the melted QGP state is in general 
entropy richer than the frozen HG phase \cite{entropy}. 
Once entropy has been generated, it cannot be lost, an entropy 
excess accompanies QGP formation. When abundant particle production 
is possible, this entropy excess is seen as an enhancement in the total 
hadronic particle multiplicity with  each 
(relativistic) meson carrying about 4 units of entropy
out of the interaction region.

Similarly, strangeness is in general more abundant 
in QGP than HG phase \cite{Raf82}, and 
it is not reannihilated in rapid decomposition of the 
dense matter state \cite{KMR86}. 
It has become a key  diagnostic tool of dense hadronic 
matter because:\\
\indent {1)} particles containing  strangeness are produced more 
abundantly in relativistic nuclear collisions than it could be 
expected based on simple scaling of $p$--$p$ reactions;\\ 
\indent {2)} all strange quarks  have to be
made, while light $u$, $d$  quarks are  also
brought into the reaction by the colliding nuclei;\\ 
\indent {3)}
because there are many different strange particles, we have a
very rich field of observables with which it is possible to explore
diverse properties of the source;\\ 
\indent {4)} theoretical calculations suggest that glue--glue
collisions in the QGP provide a sufficiently fast and
thus by far, a unique mechanism leading to an explanation of
strangeness enhancement. 

There are two generic flavor 
 observable (strangeness and charm) which we  study  
analyzing experimental data, and we introduce these here,
without an effort to `orthogonalize', {\it i.e.}, make 
them independent of each other:
\begin{itemize}
\item {absolute yield} of strangeness/charm\\ 
Once produced in hot and dense hadronic matter, {\it e.g.}, 
the QGP phase, strangeness/charm is
not reannihilated in the evolution of the deconfined state towards
freeze-out, because in the expansion and/or cooling process the rate of 
production/annihilation rapidly diminishes and becomes negligible. 
Therefore the flavor yield is characteristic of the
initial, most extreme conditions. 
 \item {phase space occupancy  $\gamma_{i}$}\\
$\gamma_{i}$ describes how close the flavor  yield per 
unit of volume ($i={\rm s,c}$) comes to the chemical equilibrium expected; 
$\gamma_i$  impacts strongly the distribution of 
flavor among final state  hadronic particles.
\end{itemize}
Because of the high density of the QGP phase, the 
phase space occupancy $\gamma_{i}$ 
can saturate rapidly, and thus particle abundances 
will emerge from a chemically equilibrated $u$, $d$, $s$ phase, which is
hardly imaginable for conventional reaction mechanisms.
Because entropy and
strangeness are enhanced in a similar way in QGP, the specific yield of
strangeness per particle produced is not a good quantity to
use when searching for the deconfined state. Many other strategies
are available, of which we favor measurement of the specific 
entropy yield per participating baryon, accompanied 
by a study of relative 
strange antibaryon yields, involving particle ratios such 
as $\overline{\Lambda}/\bar p$ \cite{Raf82}. It is remarkable that
the pertinent results obtained for S--Pb collisions 
by the NA35 collaboration \cite{NA35pb} have 
shown the QGP related enhancement. 

We use here the framework of 
local thermal equilibrium, and thus we will now briefly survey 
 the different equilibria possibly present in 
the nucleus--nucleus high energy collisions. The principal reason 
to use the statistical description in study of dynamics of dense
hadronic matter  is that considerable simplicity can 
result, compared for example with dynamical models involving 
 collisional cascading of individual particles. 
We realize considerable economy in the 
description of the physical system by invoking a local thermal 
equilibrium. When we speak of equilibrium in a rapidly evolving system, 
such as is the dense, highly excited  phase of hadronic matter,
we always imply {\it local} equilibrium: specifically, within the local
space-time region the temperature parameter $T$
describes  the energy content of the individual degrees of freedom.
Even with this restriction, we  cannot generally subsume
that statistical description always makes good sense, since 
it implies presence of many degrees of freedom in mutual interaction.
We believe that strongly interacting  degrees of freedom 
involved in collisions of large atomic nuclei will
participate in the equipartition of energy through elastic or
inelastic collisions, justifying the introduction of 
local temperature.

Since any local thermal equilibrium will be overlaid by 
diverse flow phenomena, how can we experimentally see that 
our hypothesis is indeed consistent with the data? First, we
note that spectra of particles produced have, in the transverse
mass variable $m_\bot=\sqrt{m^2+p_\bot^2}$, 
a `thermal' shape,  consistent with 
a thermal and expanding source. The inverse slope of the $m_\bot$ 
spectra provides a measure of the temperature, and it can vary 
from one kind of particle to another due to different freeze-out 
properties of different particle and mass dependence of 
flow. What supports strongly the notion of local thermal equilibrium is the
experimental fact that, at given $m_\bot$,  
very different particles have, in the same 
reaction, $m_\bot$-spectra that sport the 
same inverse slope (temperature). There seems to be presently 
no exception from this $m_\bot$-scaling rule, even though widely different 
`temperatures' are sometimes quoted for different particles; it turns
out on closer inspection that the domains of good, precise data points 
which weight heavily in the spectral fits are in very different ranges
of $m_\bot$ where different slopes are reported.
 It is thus quite difficult to compare different particles 
in the same $m_\bot$-range, with the exception of comparing particles
with antiparticles, where indeed same spectral shapes are typically
seen (see, {\it e.g.}, $\overline{\Lambda}$ and $\Lambda$ spectra).
One must
 be further aware that, even in most central symmetric collisions, a 
good fraction of colliding nucleons, 
${\cal O}(15\%)$, will not participate in the 
dynamics of centrally colliding matter, 
contributing to a longitudinal flow background.

We further recognize  two different forms of chemical equilibria.
To see the need for this differentiation 
consider a hot, thermally equilibrated
gas of colliding nucleons $N$, pions $\pi$, and $\Delta$-resonances. 
The relative abundance of the $u$ and $d$ carrying quarks is easily 
established through flavor exchange reactions, such as $p+\pi^-
\leftrightarrow n+\pi^0$, at the quark level there is no creation
or annihilation process that need to occur in each reaction to 
equilibrate the flavors. We speak in such type of reactions of 
relative chemical equilibration
through quark exchange reactions. Very much different is the approach
to equilibrium involving the equilibration of the number of 
mesons and baryons.  Baryon-antibaryon formation 
processes involving heavy mesons, that is of the type
$N+\bar N\leftrightarrow\rho+\omega$ require that aside of the 
reorganization of the quark content, also the number of valance quarks
changes.  Such processes are typically much slower than those
leading to establishment of relative chemical equilibrium, since 
the inelastic reactions that change the particle abundance have
usually much smaller cross section than particle number preserving 
exchange reaction. While thermal and relative chemical equilibrium 
will in general occur within the life span of hadronic reaction, 
approach to absolute chemical equilibrium provides interesting 
chronometric  information about the dynamics of the collision 
process. 

At the level of elementary QCD degrees of freedom,
 it is the production
of heavy flavors which is the primary off-equilibrium aspect of chemical 
equilibration. At current SPS energies, where the inverse $m_{\bot}$-slopes
observed are in the range $T=250$--$300$ MeV for strange baryons and 
other heavy particles, the chances for thermal production of charm are not
big; however, at RHIC and LHC considerable greater temperatures and possibly 
longer life span of the dense matter is expected, and thus we can hope and expect 
that it will become possible to study charm approach to absolute chemical
equilibrium.

In the next section  we survey  the  experimental 
strange particle results  obtained in Pb--Pb collisions.  
This is followed by updates of our work of last years 
\cite{acta96}: recent advance in 
developing more precise  understanding of strange and charm 
flavor production in deconfined phase is presented in 
section 3; our analysis of the  Pb--Pb  results is presented in section 4;
 in the final section we consider the possibility that quark gluon 
plasma has been discovered in the SPS experiments.

\section{Strange particles in Pb--Pb reactions}
The possibility that strange particle anomalies seen in recent
years at SPS  in Sulphur  induced reactions on heavy
nuclei are  arising in consequence to the formation of a 
deconfined QGP phase
has stimulated the intense continuation of the experimental 
research program in the considerably more difficult, 
high particle  multiplicity environment arising in Pb induced 
reactions, which are presently possible at 158A GeV.

We address \cite{fit97} several (7 today) available
ratios of strange and anti strange baryons 
measured by the WA97--collaboration \cite{WA97} and 
two such ratios measured by the NA49--collaboration \cite{NA49B,NA49O}. 
These ratios appear only on first sight to be much like the
earlier 200A GeV S--W and S--S data of experiments WA85 and WA94 
\cite{WA85}. The NA49--collaboration  has also
presented the rapidity and transverse mass spectra of $\Lambda$, 
$\overline{\Lambda}$ and kaons \cite{NA49B}; the latter result 
are consistent with the 
central transverse mass spectra already reported
by the NA44--collaboration \cite{NA44}. 
The NA49--collaboration stresses that there is no major change in 
the yield of strangeness  per particle multiplicity, 
compared to the earlier study of 
S--S and S--Au interactions \cite{NA35}; this is in agreement 
with our expectation that strangeness and entropy (multiplicity) 
grow at the same rate as we change from S- to Pb-induced reactions,
assuming that in both cases QGP/gluon fusion mechanisms (see 
next section) are at the origin of strangeness production. 

\begin{figure}[tb]
\vspace*{1.9cm}
\centerline{\hspace*{1.6cm}
\psfig{width=12cm,figure=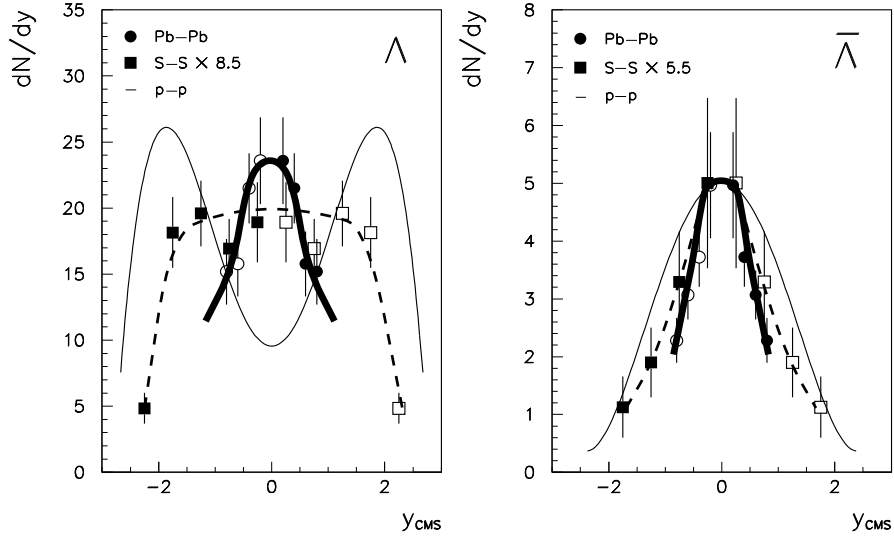}
}
\vspace*{-2.7cm}
\caption{
Rapidity distribution of $\Lambda$ (left hand side) and $\overline{\Lambda}$
(right hand side). Circles (and thick solid line to guide the eye):
 Pb--Pb collisions at 158A GeV; squares (and  dashed line) scaled 
S--S collisions at 200A GeV
(factor 8.5 for $\Lambda$ and factor 5.5 for $\overline{\Lambda}$), 
and thin solid line, shape expected from
p--p data. Open symbols: reflection around central rapidity.
Data by NA49  collaboration \protect\cite{NA49B}) \label{NA49LAL}
}
\end{figure}

First NA49 results about the production of $\phi=s\bar s$ are also 
already available \cite{NA49F}. The revised inverse slopes of the 
$\phi$ are reported at $T=280$ MeV, in agreement with temperatures of
all heavy $(m\simeq 1$ GeV) strange particles such as 
$\Lambda,\,\overline{\Lambda}$.  The experiment NA49 further 
reports  an important change in the shape of rapidity
spectra \cite{NA49B}: the $\Lambda, \overline{\Lambda}$ and  kaon 
rapidity  distributions are localized  around 
mid-rapidity and are, in particular in case of $\Lambda$, much 
narrower than previously seen in 
S-induced reactions, with a rapidity shape corresponding
to a  radiating, thermal source undergoing transverse 
and longitudinal expansion. The rapidity spectra of
$\Lambda,\,\overline{\Lambda}$ are presented in Fig.\,\ref{NA49LAL};
the scaling of the yields 
(factor 8.5 for $\Lambda$ and factor 5.5 for $\overline{\Lambda}$)
hides somewhat from view the strong concentration in Pb--Pb reactions
of the $\Lambda$ yield at central rapidity.

The WA97--collaboration stresses that an  enhancement in the 
$\Omega/\Xi$ ratio in Pb--Pb compared to p--Pb collisions:
$$
\frac{(\Omega+\overline{\Omega})/(\Xi+\overline{\Xi})|_{\rm Pb-Pb}}
     {(\Omega+\overline{\Omega})/(\Xi+\overline{\Xi})|_{\rm p-Pb\ }}
                       \simeq 3 \ \ (>2,\ \mbox{at 95\% C.L.})\,.
$$
The significance of the last result is that hadronic cascades tend to 
attenuate the yield of multistrange hadrons in  strangeness
exchange reactions and thus such an enhancement is not natural
for a reaction picture involving hadrons-hadrons collisions. 
This can be thus taken as evidence that
direct formation and emission from a hot fireball of deconfined matter is the
prevailing mechanism for the production of multistrange baryons and antibaryons. 

\begin{figure}[tb]
\vspace*{5.7cm}
\centerline{\hspace*{0.6cm}
\psfig{width=15cm,figure=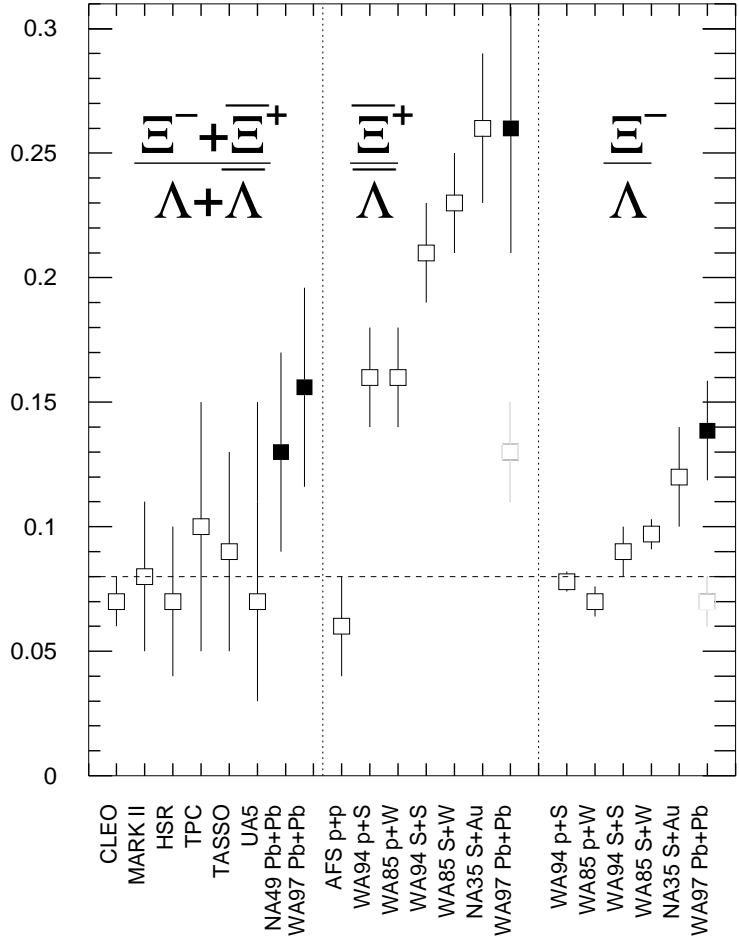}
}
\vspace*{-3.0cm}
\caption{
Sample of World results (as `function' of experiment name
for yields involving ratio of
strange to non-strange quarks in baryons.
 Dark squares: recent Pb--Pb results.
\label{RSS}
}
\end{figure}

The experiment WA97 \cite{WA97}  has further reported several specific 
strange baryon and  antibaryon  ratios from 
Pb--Pb collisions at 158A GeV, comprising 30\% of inelastic 
interactions. All ratios are obtained in an overlapping kinematic 
window corresponding effectively to
transverse momentum $ p_\bot>0.7$ GeV, within the central 
rapidity region $ y\in y_{cm}\pm0.5$. They have been corrected for weak
interactions cascading  decays. The experimental values are:
\begin{eqnarray}\label{rat1}
R_\Lambda\!=\!\frac{\overline{\Lambda}}\Lambda\! =\! 0.14\pm 0.03, \ 
R_\Xi\!=\!\frac{\overline{\Xi}}\Xi        \! =\! 0.27\pm 0.0,\ 
R_\Omega\!=\!\frac{\overline{\Omega}}\Omega  \! =\! 0.42\pm 0.12,\ \ 
\end{eqnarray}
\vspace*{-0.6cm}
\begin{eqnarray}
R_{\rm s}^{\rm p}=\frac\Xi\Lambda=0.14\pm 0.02\,,  &&\quad   
R_{\bar {\rm s}}^{\rm p} =\frac{\overline\Xi}{\overline\Lambda} = 0.26\pm 0.05\,,\\
\ \nonumber \\
{R'_{\rm s}}^{\rm p}=\frac\Omega\Xi =0.19\pm 0.04\,,  &&\quad
{R'_{\bar {\rm s}}}^{\rm p}= \frac{\overline\Omega}{\overline\Xi}  = 0.30\pm 0.09\,.
\end{eqnarray}
Here, the lower index s,
resp. $\bar {\rm s}$, reminds us that the ratio measures the density of
strange, resp. antistrange, quarks relatively to light quarks.
The upper index $p$ indicates that the ratio is taken
within a common interval of transverse momenta (and not common
transverse mass). We compare $R_{\rm s}^{\rm p},\, R_{\bar {\rm s}}^{\rm p}$   results with 
earlier measurements in Fig.\,\ref{RSS}. The strange antibaryon enhancement
effect is reconfirmed in the Pb--Pb data, and we see that there is no major 
change of this result, which determines the phase space occupancy 
of strangeness, as we move from S--S  or S--W/Pb results to Pb--Pb 
results. 

There is agreement between WA97 and NA49 
on the value of $R_\Lambda$,
even though the data sample of NA49 is taken for more central trigger,
constrained to as few as 4\% of most central  collisions. 
The cuts in $p_\bot$ and $ y$
are nearly identical in both experiments. From Fig.\,3 in 
\cite{NA49B}, we obtain the value 
$R_\Lambda=0.17\pm0.03$, which we shall combine with the value given by
WA97 and we thus take in out data fit:
\begin{eqnarray}\label{rat2}
R_\Lambda=\frac{\overline{\Lambda}}\Lambda = 0.155\pm 0.04\,.
\end{eqnarray}

We should keep in mind
that for seven different particle yields measured 
only five ratios are independent of each other. The two  constraints are:
\begin{eqnarray}
\frac{R_\Xi}{R_\Lambda}=\frac{R_{\bar {\rm s}}^{\rm p}}{R_{\rm s}^{\rm p}}\,,\qquad
\frac{R_\Omega}{R_\Xi}=\frac{{R'_{\bar {\rm s}}}^{\rm p}}{{R'_{\rm s}}^{\rm p}}\,.
\end{eqnarray}
Despite this, in our  fit  to the above data, we shall retain all seven 
independently determined ratios as presented here
and the error $\xi^2$ will correspond to sum of seven relative 
square errors. 

The experiment NA49 also reported \cite{NA49O}:
\begin{eqnarray}
\frac{\Xi+\overline{\Xi}}{\Lambda+\overline{\Lambda}}=0.13\pm0.03\,.
\end{eqnarray}
While this ratio can be expressed in terms of the three other ratios
\begin{eqnarray}
\frac{\Xi+\overline{\Xi}}{\Lambda+\overline{\Lambda}}=
R_{\rm s}\frac{1+R_\Xi}{1+R_\Lambda}\,,
\end{eqnarray}
it is again a separate measurement which thus can be fitted independently.

We will now estimate the ratio of hyperons to kaons, required 
in some of our studies below.  We note that the NA49 spectra
\cite{NA49B} of kaons and hyperons have a slightly overlapping domain of 
$m_\bot$. We recall that the slopes are 
not exactly equal, thus all we can do is to
try to combine the two  shapes, assuming continuity consistent with flow, 
and to estimate the relative normalization of both that would place
all experimental points on a common curve. We have carried out this procedure 
and obtained:
\begin{eqnarray}\label{klratio}
\left.\frac{\Lambda}{K^0_{\rm s}}\right|_{m_\bot}\simeq 6.2\pm1.5\,.
\end{eqnarray}
Note that there is a tacit presumption in our approach 
that a similar effective  $\Delta y$ interval
was used in both spectra.  We recall that this ratio
was $4.5\pm0.2$ in the S-W data \cite{WA85K}.

We now consider at the experimental inverse $m_\bot$  slopes.
In the common $p_\bot$ range of WA97 and NA49 experiments the transverse 
mass spectrum of $\Lambda$ and $\overline\Lambda$ obtained by NA49 is
very well describe by an exponential \cite{NA49B}. 
A thermal model motivated fit of the inverse slope 
(temperature) yields $T_\bot^\Lambda=284\pm15$ MeV and 
$T_\bot^{\overline{\Lambda}}=282\pm20$ MeV. This is consistent with
the mid-rapidity proton and antiproton slope of the NA44 experiment: 
$T_p=289\pm7$ MeV and $T_{\overline{p}}=278\pm9$ MeV. For $\Xi+\overline{\Xi}$
a consistent value $T_\Xi=290$ MeV is also quoted by the NA49--collaboration~\cite{NA49O}. 
We note that because the baryon masses are large, all 
these slopes are at relatively
high $m_\bot>1.3 $ GeV (for nucleons, in NA44, $m_\bot>1 $ GeV). 
Systematically smaller inverse-transverse slopes    
are reported at smaller $m_\bot$, for kaons $T_\bot^K\simeq 213$--$224$ MeV 
for $0.7 <m_\bot<1.6$ GeV in NA49 \cite{NA49B} and 
$T_\bot^{K^+}= 234\pm6,\,T_\bot^{K^-}=235\pm7$ MeV in NA44 \cite{NA44};
and 155--185 MeV for $\pi$, \cite{NA49B,NA44},  depending 
on the range of $p_\bot$, but here we have to remember that pions
are known to arise primarily from resonance decays. 
An increase of $T$ with $m_\bot$ 
 is most naturally associated with the effects of transverse flow of
the source.

\section{Strangeness Production in QGP}
\subsection{Thermal reaction rates}
We use  two particle collision 
processes  to evaluate thermal flavor 
production in QGP, as described in section 6 in Ref.\,\cite{acta96},
see also Ref.\,\cite{impact96}.
Recent results about strong interaction reaction rates constrain
to sufficient precision the coupling strength $\alpha_{\rm s}(\mu)$,
however  considerable
uncertainty still remains, in particular in regard of strange quark
mass scale $m_{\rm s}(\mu)$, as well as systematic uncertainty related to applications
of QCD to soft (less than 1 GeV) processes. 

To determine these two QCD parameters ($\alpha_{\rm s}$ and $m_{\rm s}(\mu)$), 
we will use the
renormalization group functions $\beta$ and $\gamma_{\rm m}$:
\begin{equation}\label{dmuda}  
\hspace*{-0.5cm}\mu \frac{\partial\alpha_{\rm s}}{\partial\mu}
=\beta(\alpha_{\rm s}(\mu))\,,\qquad
\mu {\frac{\partial m}{\partial\mu}} =-m\,
\gamma_{\rm m}(\alpha_{\rm s}(\mu))\,.\hspace*{-0.5cm}
\end{equation}
We use the perturbative power expansion 
in $\alpha_{\rm s} $:
\begin{eqnarray}\label{betaf}
\beta^{\rm pert}&=&\alpha_{\rm s}^2\left[\ b_0
   +b_1\alpha_{\rm s} +b_2\alpha_{\rm s}^2 +\ldots\ \right] \,,
\nonumber\\ 
\label{gamrun}
\gamma_{\rm m}^{\rm pert}&=&\alpha_{\rm s}\left[\ c_0
+c_1\alpha_{\rm s} +c_2\alpha_{\rm s}^2 + \ldots\ \right]\,,
\end{eqnarray}
For the SU(3)-gauge theory with $n_{\rm f}$ fermions the first two
terms are renormalization scheme independent, and
we include in our calculations the next three `loop' term as well, which
is  renormalization scheme dependent, evaluated in the 
MS-scheme \cite{SS96}. 
\begin{eqnarray}
b_0&=&
\displaystyle\frac{1}{ 2\pi}
  \left(11-{2\over 3}n_{\rm f}\right),\hspace*{0.5cm}
b_1=\,\frac{1}{4\pi^2}\left(51-{19\over 3}
        n_{\rm f}\right), \\
b_2&=&
\displaystyle\frac{1}{64\pi^3}\left(2857-{\frac {5033}{9}}\,n_{\rm f}
+{\frac {325}{27}}\,{n_{\rm f}}^{2}\right),\ \nonumber\\
c_0&=&
 \displaystyle\frac{2}{\pi},\hspace*{0.5cm}  
c_1=\frac{1}{12\pi^2}
        \left(101-{10\over 3}n_{\rm f}\right),\\
c_2&=&
\displaystyle\frac{1}{32\pi^3}\left(1249
-\left({2216\over 27}+{160\over 3}\zeta(3)\right)n_{\rm f}
-{140\over 81}{n_{\rm f}}^{2}\right).\nonumber
\end{eqnarray}
The numerical influence of the third order term is 
nearly negligible for $\mu> 1$ GeV, and it serves to stabilize 
our numerical solutions in the infrared domain. For this reason, we
did not introduce the scheme dependent cross section to make the
physical process rates exactly calculation scheme independent.

\begin{figure}[tb]
\vspace*{-1.1cm}
\centerline{\hspace*{0.6cm}
\psfig{width=11cm,figure=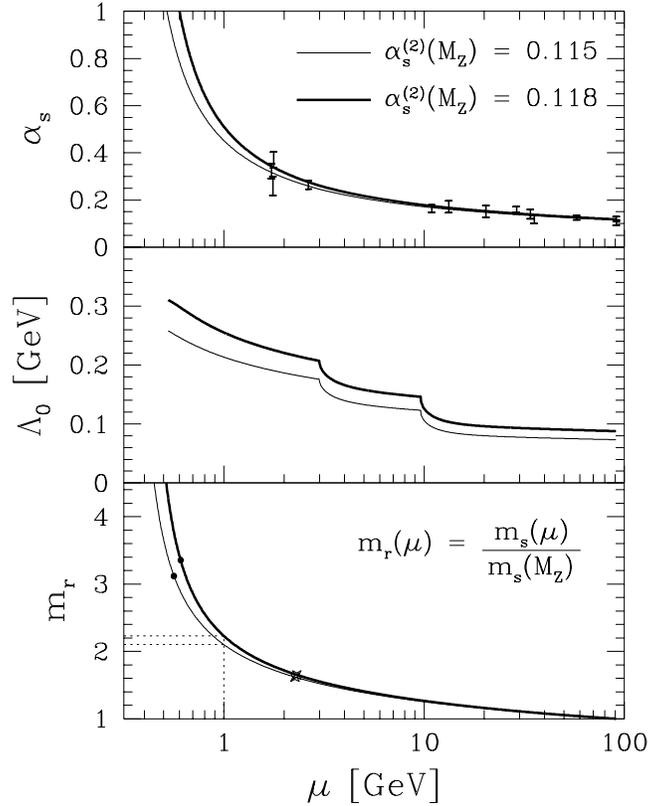}
}
\vspace*{0.6cm}
\caption{
\rm$\alpha_{\rm s}(\mu)$ (top section); the equivalent 
parameter $\Lambda_0$ (see Eq.\,\protect\ref{Lambdarun}, middle section) and
$m_{\rm r}(\mu)=m(\mu)/m(M_{Z})$
(bottom section) as function of energy scale $\mu$. 
Initial value $\alpha_{\rm s}(M_{Z})=0.118$ (thick solid lines)
and $\alpha_{\rm s}(M_{Z})=0.115$ (thin solid lines).
In lower section the dots indicate the strangeness pair production
thresholds for $m_{\rm s}(M_{Z})=$~90~MeV, while crosses indicate
charm pair production thresholds for $m_{\rm c}(M_{Z})=$~700~MeV.
\label{fig-a1}}
\end{figure}

We use, in this report,
the August 1996 World average \cite{Sch96}: 
$\alpha_{\rm s}(M_{{Z}})=0.118$ for which the estimated error is 
$\pm\ 0.003$\,. This value is sufficiently precise to eliminate most
of the uncertainty that has befallen much of our earlier
studies \cite{acta96,impact96}.
First equation in
(\ref{dmuda}) is numerically integrated  beginning with an initial
value of $\alpha_{\rm s}(M_Z)=0.118$ and result is shown in the 
top portion of  Fig.\,\ref{fig-a1} (thick solid line). 
The thin solid line
present results for $\alpha_{\rm s}(M_{{Z}})=0.115$\, till recently
the preferred result in some analysis, especially those at lower
energy scale. The variation of
$\alpha_{\rm s}$ with the energy scale is substantial, and in
particular we note the rapid change at and below $\mu=1$ GeV, where
the strange quark flavor formation occurs in hot QGP phase formed
in present day experiments at 160--200A GeV (SPS). Clearly,
use of constant value of $\alpha_{\rm s}$ is hardly justified, and
the first order approximation often used
\begin{equation}\label{Lambdarun}
\alpha_{\rm s}(\mu)\equiv
  \frac{2b_0^{-1}(n_{\rm f})}{\ln(\mu/\Lambda_0(\mu))^2}\,,
\end{equation}
leads to a strongly scale dependent $\Lambda_0(\mu)$ shown in the
middle  section of  Fig.\,\ref{fig-a1}. 

With $\alpha_{\rm s}(\mu)$ from the solutions described above, we
integrate the running of the quark masses, the second  equation in
(\ref{dmuda}).  Because the running mass equation 
is linear in $m$, it is possible to determine the universal 
quark mass scale factor
\begin{equation}
m_{\rm r}=m(\mu)/m(\mu_0)\,.
\end{equation}
Since  $\alpha_{\rm s}$ refers to  the scale of $\mu_0=
M_Z$, it is a convenient reference point also for quark masses.  
As seen in the bottom portion of Fig.\,\ref{fig-a1},
the change in the quark mass factor is highly relevant,
since it is driven by the rapidly changing
$\alpha_{\rm s}$ near to $\mu\simeq 1$~GeV.
For each of the different functional dependences
$\alpha_{\rm s}(\mu)$, we obtain a different function
$m_{\rm r}$.  
Given these results, we find that for $\alpha_{\rm s}=0.118$ and
$m_{\rm s}(M_{{Z}})=90\pm18$~MeV a low energy strange quark mass
$m_{\rm s}(1\mbox{\,GeV})\simeq 200\pm 40$ MeV, in the middle of
the standard range $100<m_{\rm s}$(1\,GeV) $<$ 300 MeV. Similarly we
consider $m_{\rm c}(M_{{Z}})=700\pm50$~MeV, for which value 
we find  the low energy mass $m_{\rm c}(1\mbox{\,GeV})\simeq 1550\pm110$
MeV, at the upper (conservative for particle production yield) end
of the standard range $1<m_{\rm c}$(1\,GeV) $<1.6$ MeV. 

There is another
(nonperturbative) effect of running mass, related to the mass at
threshold for pair production $m^{\rm th}_i,\,i=$ s, c, arising from 
the solution of:
\begin{equation}\label{dispersion} 
m_i^{\rm th}/m_i(M_{{Z}})= m_{\rm r}(2m_i^{\rm th})\,.
\end{equation}
This effect stabilizes strangeness production cross section in the
infrared: below $\sqrt{\rm s}=1$ GeV the strange quark mass
increases rapidly and the threshold mass 
is considerably greater than  $m_{\rm s}$(1 GeV).
We obtain the threshold values $2m_{\rm s}^{\rm th}=611$ MeV
for  $\alpha_{\rm s}(M_Z)=0.118$ 
and $2m_{\rm s}^{\rm th}=566$ MeV for $\alpha_{\rm s}(M_Z)=0.115$. 
Both values are indicated by the black dots in 
Fig.\,\ref{fig-a1}.  For charm, the running mass effect plays
differently:  since the mass of charmed quarks is listed in tables
for $\mu=1$ GeV, but the value of the  mass is above 1 GeV,  the
production threshold mass is smaller than the listed
value $m_{\rm c}$(1 GeV). 
For $m_{\rm c}(M_Z)= 700$ MeV the
production threshold is found at $\sim 2m_{\rm c}^{\rm th}\simeq 2.3$ GeV
rather than 3.1 GeV that would have been naively expected. 
This reduction in threshold enhances thermal production
of charm, especially so at low temperatures.
The significance of the running of the charmed
quark mass cannot be stressed enough, especially for thermal charm
production occurring in foreseeable future experiments well below
threshold, which amplifies the importance of exact value of 
$m_{\rm c}$\,.

The thermal production rates are obtained from thermal average
of the cross section:
\begin{eqnarray}
A_{\rm s}&\equiv&
A_{\rm gg}+A_{\rm u\bar u}+A_{\rm d\bar
d}+\ldots\nonumber\\
&=&
\int_{4m_{\rm s}^2}^{\infty}ds 2s\delta (s-(p_1+p_2)^2)
\int{d^3p_1\over2(2\pi)^3E_1}
\int{d^3p_2\over2(2\pi)^3 E_2}     \label{qgpA}\\
&&\times
\left[{1\over 2} g_{\rm g}^2f_{\rm g}(p_1)f_{\rm g}(p_2)
\overline{\sigma_{\rm gg}}(s)       
+ n_{\rm f}g_{\rm q}^2 f_{\rm q}(p_1) 
f_{\bar {\rm q}}(p_2)\overline{\sigma_{{\rm q}\bar
{\rm q}}}(s)+\ldots\vphantom{{1\over 2}}\right]\!.\nonumber
\end{eqnarray}
The dots indicate that other mechanisms may contribute to
strangeness production. The particle distributions $f_i$ are in our
case thermal Bose/Fermi functions (for fermions with 
$\lambda_{\rm q}=1.5$), and $g_{\rm q}=6,\,g_{\rm g}=16$\,. For strangeness
production $n_{\rm f}=2$, and for charm production $n_{\rm f}=3$\,.
The generic angle averaged two particle cross section 
for (heavy) flavor production processes
\mbox{$g+g\to f+\bar f $ and $ q+\bar q\to f+\bar f\,,$ are}
\begin{eqnarray}
\bar\sigma_{\rm gg\to f\bar f}\!&=&\!
   {2\pi\alpha_{\rm s}^2\over 3s} \!\left[  \!
\left( 1\! +\! {4m_{\rm f}^2\over s}\! + \!{m_{\rm f}^4\over s^2} \right) 
\! {\rm tanh}^{-1}W(s)\!-\!\left({7\over 8}\! + \!{31m_{\rm f}^2\over
8s}\right)\! W(s) \right]\!,\ \ \ \\
\bar\sigma_{{\rm q}\bar {\rm q}\to f\bar f} \!&=&\!
   {8\pi\alpha_{\rm s}^2\over 27s}
   \left(1+ {2m_{\rm f}^2\over s} \right) W(s)\,,
\label{gk}
\end{eqnarray}
where $W(s) = \sqrt{1 - 4m_{\rm f}^2/s}$\,, and both the QCD
coupling constant $\alpha_{\rm s}$  and flavor quark mass $m_{\rm f}$ 
are as determined above. 

\begin{figure}[ptb]
\vspace*{-0.6cm}
\centerline{\hspace*{-0.5cm}
\psfig{width=11.5cm,figure=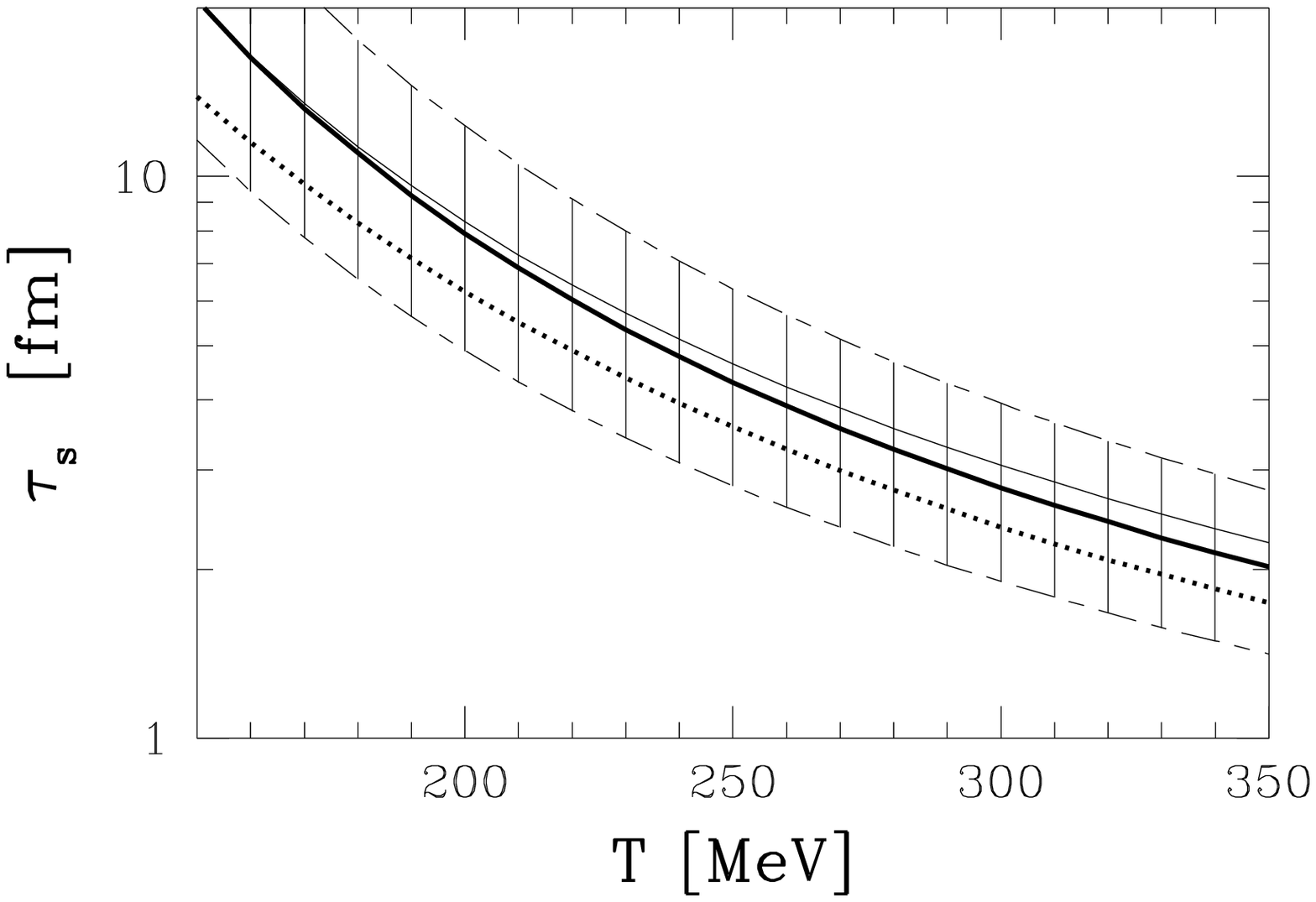}}
\vspace*{0.4cm}
\caption{
QGP strangeness relaxation time, for $\alpha_{\rm s}(M_{Z})=0.118$,
(thick line) and = 0.115 (thin line); $m_{\rm s}(M_{{Z}})=90$~MeV.
Hatched areas: effect of variation of strange quark mass by 20\%.
 Dotted: comparison results for fixed  
$\alpha_{\rm s}=0.5$ and $m_{\rm s}=200$ MeV. 
}
\label{figTaussrun}
\vspace*{0.6cm}
\centerline{\hspace*{-0.5cm}
\psfig{width=11.5cm,figure=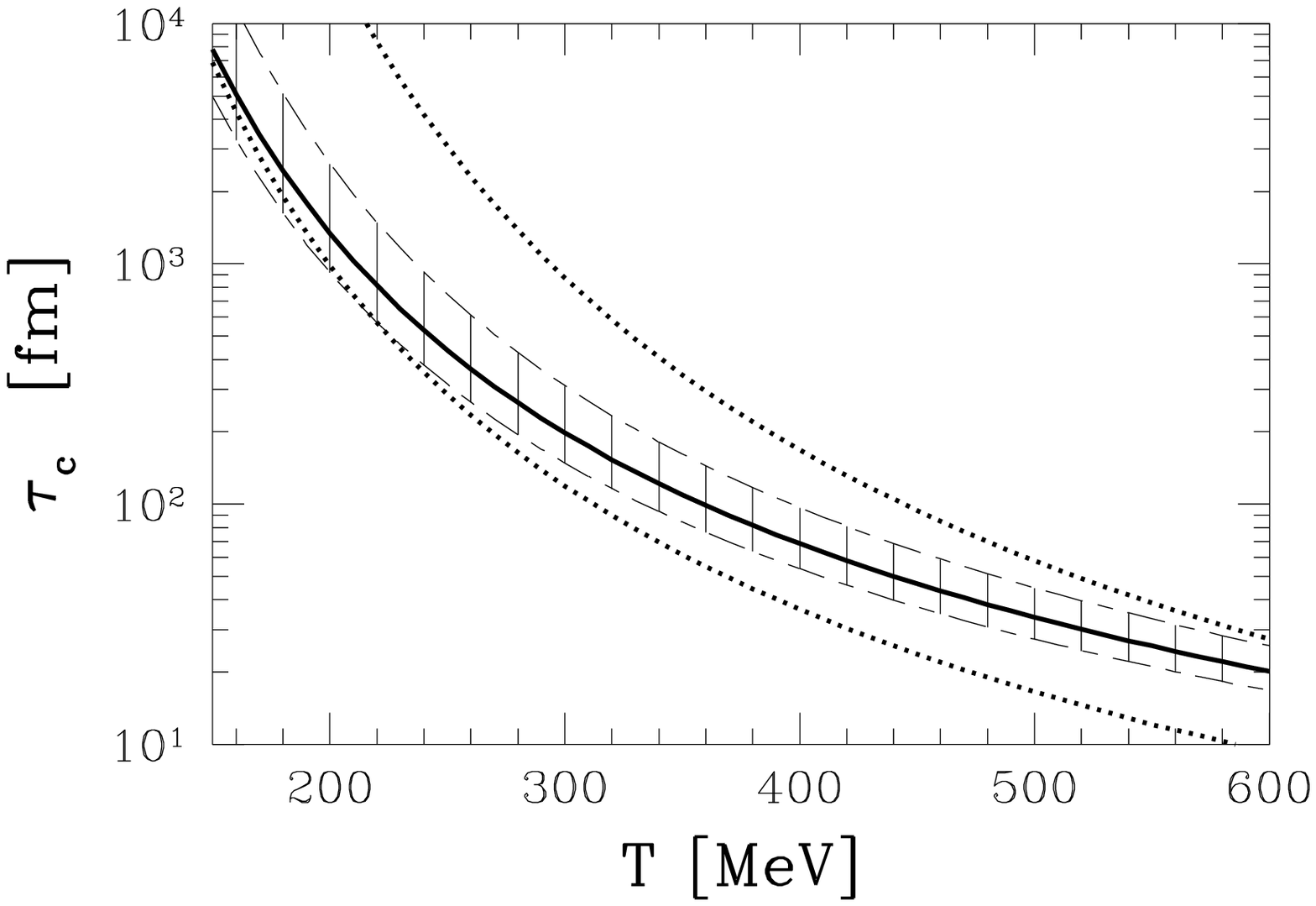}
}
\vspace*{0.4cm}
\caption{
Solid lines: thermal charm relaxation constant in QGP, calculated
for running  $\alpha_{\rm s}(M_{Z})=$ 0.115; 0.118, (indistinguishable),
$m_{\rm c}(M_Z)= 700$ MeV. Lower dotted line: for fixed 
$m_{\rm c}=1.1$ GeV, $\alpha_{\rm s}=0.35$;
upper doted  line: for fixed $m_{\rm c}=1.5$ GeV, $\alpha_{\rm s}=0.4$\,.
Hatched area: effect of variation $m_{\rm c}(M_Z)= 700\pm 50$~MeV
\label{figtaucc}
}
\vspace*{-1cm}
\end{figure}

From the invariant rate we obtain the strangeness relaxation time
$\tau_{\rm s}$ shown in Fig.\,\ref{figTaussrun}, as function of
temperature:
\begin{equation}\label{tauss}
\tau_{\rm s}\equiv
{1\over 2}{\rho_{\rm s}^\infty(\tilde m_{\rm s})
\over{(A_{\rm gg}+A_{\rm qq}+\ldots)}}\,. 
\end{equation}
Note that here yet unaccounted for processes, such as the 
 odd-order in $\alpha_{\rm s}$ would add to the production
rate incoherently, since they can be distinguished by the presence
of incoming/outgoing gluons. Thus the current calculation offers an
upper limit on the actual relaxation time, which may still be
smaller. In any case,  the present result suffices to confirm
that strangeness will be very near to chemical equilibrium in QGP
formed in collisions of large nuclei. 
 
We  show in Fig.\,\ref{figTaussrun} 
the impact of a 20\% uncertainty in $m_{\rm s}(M_{{Z}})$, 
indicated by the hatched areas. This uncertainty is today
much larger compared to the uncertainty that arises from the
recently improved precision of the strong coupling constant
determination \cite{Sch96}.  We note that
the calculations made at fixed values 
$\alpha_{\rm s}=0.5$ and $m_{\rm s}=200$~MeV \cite{sprodQGP} 
(dotted line in Fig.\,\ref{figTaussrun}) are well within the band
of values related to the uncertainty in the strange quark mass.

Since charm is somewhat more massive compared to strangeness, there
is still less uncertainty arising in the extrapolation of the
coupling constant. Also the systematic uncertainty related to the
soft gluons (odd-$\alpha_{\rm s}$) terms are smaller, and thus the
relaxation times $\tau_{\rm c}$, we show in Fig.\,\ref{figtaucc},
are considerably better defined compared to $\tau_{\rm s}$. There
is also  less relative uncertainty in the value of charm mass.
We show in Fig.\,\ref{figtaucc} (dotted lines) the fixed 
$m_{\rm c},\, \alpha_{\rm s}$ results with parameters selected to
border high and low $T$ limits of the results presented. It is
difficult to find a good comparative behavior of $\tau_{\rm c}$ 
using just one set of $m_{\rm c}$ and $\alpha_{\rm s}$. This may be
attributed to the importance of the mass of the charmed quarks,
considering that the threshold for charm production is well above
the average thermal collision energy, which results in emphasis of
the effect of running charm mass. In the high $T$ limit, the choice
(upper doted line in Fig.\,\ref{figtaucc}) $m_{\rm c}=1.5$ GeV,
$\alpha_{\rm s}=0.4$ is appropriate, while to follow
the result at small $T$  (lower doted line in
Fig.\,\ref{figTaussrun}) we take a much smaller 
mass $m_{\rm c}=1.1$ GeV, $\alpha_{\rm s}=0.35$\,.
 
We recall that the equilibrium distribution is result of Boltzmann 
equation description of two body collisions. Thus the mass arising
in the equilibrium density $\rho_{\rm s}^\infty$ in
Eq.\,(\ref{tauss}), is to be taken at the energy scale of the
average two parton collision. We adopt for this purpose a fixed
value ${\tilde m}_{\rm s}=200$~MeV, and observe that in the range
of temperatures here considered the precise value of the mass is
insignificant, since the quark density is primarily governed by the
$T^3$ term in this limit, with finite mass correction being ${\cal
O}$(10\%). The situation is less clear for charm relaxation, since
the running of the mass should have a significant impact. Short of
more complete kinetic treatment, we used $m_{\rm c}\simeq 1.5$ GeV
in order to establish the reference density $\rho_{\rm c}^\infty$
in Eq.\,(\ref{tauss}). 
  
\subsection{Strangeness evolution and hadronization}
Given the improved relaxation times we proceed as in 
Ref.\,\cite{acta96} to determine the two important 
strangeness observables ($\gamma_{\rm s},\ N_{\rm s}/B$) 
under conditions applicable for the SPS 
reactions. In order to compute the production and evolution of
strangeness (and charm) flavor a more specific picture of the
temporal evolution of dense matter is needed. In a simple, qualitative
description, we assume that the hot, dense matter is homogeneous. 
We consider that, in Pb--Pb collisions at SPS,
the radial expansion is the dominant
factor for the evolution of the fireball properties such as
temperature/energy density and lifetime of the QGP phase. 
The expansion dynamics follows from two assumptions: \\ 
\noindent $\bullet$ the (radial) expansion 
is entropy conserving, thus the volume and temperature satisfy
\begin{equation}\label{adiaex}
V\cdot T^3=\,{\rm Const.}
\end{equation}
\noindent $\bullet$ the surface flow velocity is given by the 
sound velocity in  a relativistic gas
\begin{equation}
v_{\rm f}=1/\sqrt{3}\,.
\end{equation}

These two assumption imply the following explicit forms 
for the radius of the fireball and its average temperature:
 \begin{equation}
R=R_{\rm in}+{1\over \sqrt{3}}(t-t_{\rm in}),\
\label{T(t)} 
T={T_{\rm in}
\over{1+({t-t_{\rm in}})/\sqrt{3}R_{\rm in}}}.\nonumber
\end{equation}
We shall see below that if QGP formation is involved, 
a fit of strange antibaryons data either leads to direct emission
before expansion, or to emission from a surface expanding with 
just this velocity $v_{\rm f}$.

The initial conditions have been determined such that the energy per baryon is 
given by energy and baryon flow, and the total baryon number is 
$\eta (A_1+A_2)$, as stopped in the interaction region.
They are shown in table~\ref{initialc}.
 The radius are for zero impact parameter.
For this, equations of state of the 
QGP are needed, and we have employed our model \cite{analyze2}  in which the
perturbative correction to the number of degrees of freedom were 
incorporated along with thermal particle masses.
 
\begin{table}[t]
\caption{\label{initialc}
The initial conditions for S--Pb/W at 200A GeV and Pb--Pb at 158A GeV
for different stopping values $\eta$.}
\begin{center} 
\begin{tabular}{l|ccccc} 
\hline\hline\vphantom{$\displaystyle\frac{E}{B}$}
&$ t_{\rm in}$ [\rm fm] & $\eta$ & $R_{\rm in}$  [\rm fm] 
& $T_{\rm in}$ [{\rm MeV}]& $\lambda_{\rm q}$\\
\hline
S--Pb/W &1&0.35&3.3&280&1.5\\
&1&0.5&3.7&280&1.5\\
\hline
Pb--Pb &1&0.5&4.5&320&1.6\\
&1&0.75&5.2&320&1.6

\end{tabular} 
\end{center} 
\vspace{-0.4cm}
\end{table} 

Allowing for dilution of the phase space density 
in expansion, we integrate a population 
equations describing the change in $\gamma_{\rm s}(t)$ \cite{acta96}:
\begin{equation}\label{dgdtf}
\hspace*{-0.2cm}{{d\gamma_{\rm s}}\over{dt}}\!=\!
\left(\!\gamma_{\rm s}{{\dot T m_{\rm s}}\over T^2}
     {d\over{dx}}\ln x^2K_2(x)\!+\!
{1\over 2\tau_{\rm s}}\left[1-\gamma_{\rm s}^2\right]\!\right).
\end{equation}
Here K$_2$ is a Bessel function and $x=m_{\rm s}/T$. 
Note that even when $1-\gamma_{\rm s}^2<1$ we still can have 
a positive derivative of $\gamma_{\rm s}$, since the first term
on the right hand side of Eq.\,(\ref{dgdtf}) is always positive,
both $\dot T$ and $d/dx(x^2K_2)$ being always negative. This shows
that dilution due to expansion effects in principle can make the
value of $\gamma_{\rm s}$ rise above unity.
 
Given  the relaxation constant $\tau_{\rm s}(T(t))$, these
equations can be integrated numerically, and we can obtain 
for the two currently explored experimental systems the values of the
two observables, $\gamma_{\rm s}$ and $N_{\rm s}/B$, which are given in 
table~\ref{yield1}.
\begin{table}[t]
\caption{\label{yield1}$\gamma_{\rm s}$ and  $N_{\rm s}/B$ 
in S--W at 200A GeV and Pb--Pb at 158A GeV
for different stopping values of baryonic number and energy 
$\eta_{\rm B}=\eta_{\rm E}$\,; computed for strange quark mass 
$m_{\rm s}(1 GeV)=200\pm40$ MeV, $\alpha_{\rm s}(M_Z)=0.118$\,.}
\begin{center} 
\begin{tabular}{l|cc|cc} 
\hline\hline\vphantom{$\displaystyle\frac{E}{B}$}
$E_{\rm lab}$&\multicolumn{2}{|c}{S--W at 200A GeV}&%
\multicolumn{2}{|c}{Pb--Pb at 158A GeV}\\
\hline
$\eta_{\rm B}=\eta_{\rm E}$&0.35&0.5&0.5&0.75  \\
$\gamma_{\rm s}$& $0.53\pm 0.14$   &$0.65\pm 0.15$ &%
$0.69\pm 0.15$ &$0.76\pm 0.16$ \\
$N_{\rm s}/B$&$0.67\pm 0.16$&$0.70\pm 0.16$&%
$0.954\pm 0.20$&$0.950\pm 0.20$
\end{tabular} 
\end{center} 
\vspace{-0.4cm}
\end{table} 
There is a considerable uncertainty due to 
the unknown mass of strange quarks. However, since this is a 
 not statistical but systematic 
uncertainty depending nearly alone on the value of the 
strange quark mass parameter, all the results presented will shift
together. We note further that there seems to be 
very little dependence on the stopping fractions in
the yield of strange quarks per baryon $N_{\rm s}/B$. Thus if 
the expected increase in  stopping is confirmed, we should
also expect a small increase by 15\% in specific strangeness yield.

\section{QGP Thermo-chemical Freeze-out Parameters From Pb--Pb}
\subsection{Primordial emission scenario}
We next introduce all the model parameters used in the
fit of the particle ratios, not all will be required in 
different discussions of the experimental data. For more details
about  the thermo-chemical parameters
we refer to the extensive discussion in the
earlier study of S--S and S--W data 
\cite{acta96}. The key parameters are:\\ 
\indent 1) $T_{\rm f}$ is the formation/emission/freeze-out temperature,
depending on the reaction model. 
$T_{\rm f}$ enters in the fit of abundance ratios of unlike particles
presented within a fixed $p_\bot$ interval. The temperature $T_{\rm f}$ 
can in first approximation be related 
to the observed high-$m_\bot$ slope $T_\bot$ by:
\begin{equation}\label{shiftT}
T_\bot \simeq T_{\rm f} \frac{1+v_\bot}{\sqrt{1-v_\bot^2-v_\parallel^2}}\,.
\end{equation}
In the central rapidity region the longitudinal 
flow $v_\parallel\simeq 0$, in order to assure
 symmetry between projectile and target. Thus as long as  $T_{\rm f}<T_\bot$, 
we shall use Eq.\,(\ref{shiftT}) setting $v_\parallel=0$ 
to estimate  the transverse 
flow velocity $v_\bot$ of the source.\\ 
\indent 2) $\lambda_{\rm q}$ is the light quark fugacity.
We initially used in our fits both $u,\,d$-flavor fugacities
$\lambda_{\rm u}$ and $\lambda_{\rm d}$, but
we saw that the results were equally adequate
without allowing for up-down quark asymmetry, using  the 
geometric average $\lambda_{\rm q}=\sqrt{\lambda_{\rm u}\lambda_{\rm d}}$; 
moreover the 
fitted up-down quark fugacity asymmetry was found
 as expected in 
our earlier analytical studies \cite{analyze1}.\\ 
\indent 3) $\lambda_{\rm s}$ is the strange quark fugacity. A source in which
the carriers of $s$ and $\bar s$ quarks are symmetric this parameter
should have the value $\lambda_{\rm s}\simeq1$, in general in a re-equilibrated 
hadronic matter the value of $\lambda_{\rm s}$ can be determined requiring
strangeness conservation.\\  
\indent 4) $\gamma_{\rm s}$ is the strange phase space occupancy. Due to 
rapid evolution of dense hadronic matter it is in general highly 
unlikely that the total abundance of strangeness can follow the 
rapid change in the conditions of the source, and thus in general 
the phase space will not be showing an overall abundance equilibrium 
corresponding to the momentary conditions.\\  
\indent 5)  We also show when appropriate  in table~\ref{t1} 
the parameter $R_{\rm C}^{\rm s}$ describing the relative 
off-equilibrium abundance of
strange mesons and baryons, using thermal equilibrium abundance as reference. 
This parameter is needed, when we have constraint on the strangeness
abundance and/or when  we address the abundance of mesons 
since there is no a priori assurance that the relative
production/emission strength of strange mesons and baryons should 
proceed according to 
relative strength expected from thermal equilibrium. Moreover, it is 
obvious that even if reequilibration of particles in hadronic gas should occur, 
this parameter will not easily find its chemical equilibrium 
value $R_{\rm C}^{\rm s}=1$
as we alluded to in section 1. However, 
due to reactions connecting strange with non-strange particles we 
expect $R_{\rm C}^{\rm s}=R_{\rm C}$, where $R_{\rm C}$ is the same ratio  for non-strange mesons 
and baryons, using thermal abundance as reference. The value of $R_{\rm C}>1$
implies meson excess abundance per baryon, and thus excess specific
entropy production, also expected in presence of
 color  deconfinement \cite{entropy}. 

The relative number of particles of same type 
emitted at a given instance by a hot source is obtained by
noting that the probability to find all the $j$-components contained within 
the $i$-th  emitted particle is
\begin{equation}\label{abund}
N_i\propto \gamma_{\rm s}^k\prod_{j\in i}\lambda_je^{-E_j/T}\,,
\end{equation}
and we note that the total energy and fugacity of the particle is:
\begin{equation}
E_i=\sum_{j\in i}E_j,\qquad \lambda_i=\prod_{j\in i}\lambda_j\,.
\end{equation}
The strangeness occupancy $\gamma_{\rm s}$ enters Eq.\,(\ref{abund})
with power $k$,  which equals the number of strange and antistrange quarks in 
the hadron $i$. 
With $E_i=\sqrt{m_i^2+p^2}=\sqrt{m_i^2+p_\bot^2}\cosh y $ 
 we integrate over the transverse momentum range 
as given by the experiment (here $p_\bot>0.6 $ GeV)
taking  central rapidity region $y\simeq 0$
to obtain the relative strengths of particles produced. 
We then allow all hadronic resonances to disintegrate 
in order to obtain the final relative multiplicity of `stable' particles
required to form the observed particle ratios. 
This approach allows to compute the relative
strengths of strange (anti)baryons both in case of 
surface emission and equilibrium disintegration of a particle gas since
the phase space occupancies are in both cases properly accounted for by 
Eq.\,(\ref{abund}). The transverse flow phenomena enter in a similar fashion into
particles of comparable mass and are not expected to
influence particle ratios. 
 Finally we note  that  particles which are easily influenced by the
medium, such as $\phi$, require a greater effort than this
simple model, and are also not explored in depth here.

{\begin{table}[t]
\caption{\small
Values of fitted statistical parameters within thermal model,
for 158A GeV Pb--Pb strange particle production data. 
Superscript star `*': a fixed input value for equilibrium 
hadronic gas;  subscript `$|c$':
value is result of the imposed strangeness conservation constraint. 
$\chi^2$ is  the total relative square 
error of the fit for  all data points used. First result line: direct emission
QGP model, no meson to baryon ratio fit. Second line: same, but with strangeness
conservation yielding $\lambda_{\rm s}$, and $R_{\rm C}^{\rm s}$ variable. 
Line three: as in line two, in addition the meson to baryon 
ratio Eq.\,(\protect\ref{klratio}) is fitted. Line four: hadronic gas fit
including the ratio Eq.\,(\protect\ref{klratio}).
\label{t1}}
\vspace{0.1cm}
\begin{center}
\begin{tabular}{l|ccccc} \hline\hline 
$T_{\rm f} [MeV]$& $\lambda_{\rm q}$&$\lambda_{\rm s}$&
$\gamma_{\rm s}$&$R_{\rm C}^{\rm s}$& $\chi^2$$\vphantom{\ds\frac{\Xi}{\Lambda}}$ \\
\hline
                   272 $\pm$ 74
                 & 1.50 $\pm$ 0.07
                 &   1.14 $\pm$ 0.04
                 &   0.63 $\pm$ 0.10
                 &   ---
                 &   1.0  \\

                   272 $\pm$ 74
                 & 1.50 $\pm$ 0.08
                 &   1.14$_{|c}$
                 &   0.63 $\pm$ 0.10
                 &   4.21$\pm$ 1.88
                 &   1.0  \\
\hline

                     151 $\pm$ 10
                 &   1.54 $\pm$ 0.08
                 &   1.13$_{|c}$
                 &   0.91$\pm$0.09
                 &   0.85$\pm$0.22
                 &     1.5      \\

                     155 $\pm$ 7\phantom{0}
                &  1.56 $\pm$ 0.09
                &  1.14$_{|c}$
                &  1$^*$
                &  1$^*$
                &   7.6   \\

\end{tabular}\\
\end{center} 
\vspace{-0.4cm} 
\end{table}}


We obtain the least square fit for the eight above reported
 (anti)baryon ratios. 
Our first approach is motivated by the reaction picture consisting of
 direct emission from the 
QGP deconfined fireball. The value of statistical parameters controlling
the abundances are thus free of constraints arising in an equilibrated
hadronic gas (HG) state \cite{analyze1}. 
The fitted thermal parameters are presented in the
first line of table~\ref{t1} along with the total $\chi^2$ for
the eight  ratios. The fit is quite good, the 
error shown corresponds to the total accumulated error from 8 measurements;
even if one argues 
that it involves 4 parameters to describe 5 truly independent quantities, 
the statistical significance is considerable, considering that 8 different 
measurements are included.
Such a free fit does not know that we are expecting that the final 
state comprises a balance $\langle s-\bar s\rangle=0$.
In order to estimate what would be implied by strangeness conservation 
constraint among emitted hadrons we present in second line of table~\ref{t1} 
the result of a fit assuming that the value of $\lambda_{\rm s}$ 
is result of the conservation constraint $\langle s-\bar s\rangle= 0$,
and allowing  $R_{\rm C}^{\rm s}\ne 1$, for there should 
be no chemical equilibrium 
among the emitted  strange mesons and strange  baryons
in a sudden and early QGP disintegration. The statistical  error is found
 the same as in line 1, since this approach substitutes one parameter
 ($\lambda_{\rm s}$) by another ($R_{\rm C}^{\rm s}$). 
The implication of this fit is that 
there must be either an excess of strange mesons or 
depletion of strange baryons
compared to thermal equilibrium expectation, since as we recall 
$R_{\rm C}^{\rm s}=C_{\rm M}^{\rm s}/C_{\rm B}^{\rm s}$ with $C_i$ 
being the yield of particles, normalized
to one for thermal equilibrium yield. 
In any case we see that strange meson excess is required, which
is consistent with excess of entropy production.

\begin{figure}[tb]
\vspace*{-1.6cm}
\centerline{\hspace*{-.8cm}
\psfig{width=8cm,figure=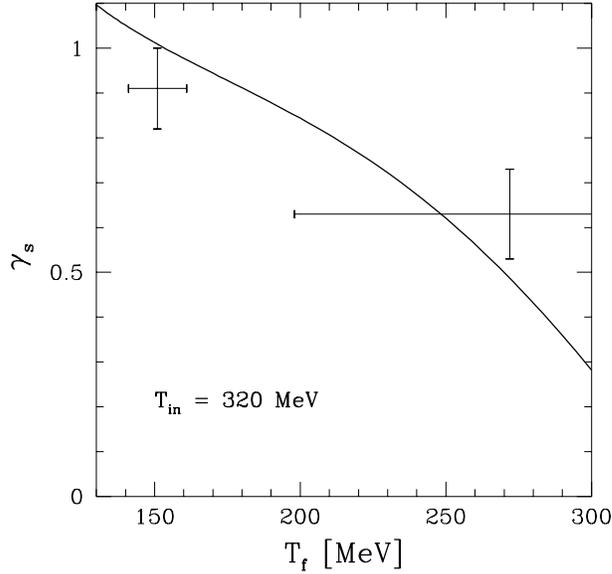}
}
\vspace*{-1.3cm}
\caption{\small
QGP strangessnes occupancy $\gamma_{\rm s}$ as function
of temperature $T_{\rm f}$ at time of particle production, for 
intial temperature $T_{\rm in}=320$ MeV, with 
$\gamma_{\rm s}(T_{\rm in})=0.1$. 
\label{gammTf}
}

\end{figure}

The errors seen in the two first lines of table~\ref{t1} 
 on the statistical parameters arise in part from strong
correlations among them. In particular
  the very large error in $T_{\rm f}$ arise from 
the 80\% anti-correlation with $\gamma_{\rm s}$.  However, some further 
information about the relation of $T_{\rm f}$ and $\gamma_{\rm s}$ 
may be garnered from theoretical considerations. We 
evaluate using our  dynamical strangeness production
model in QGP  how the
value of $\gamma_{\rm s}$ depends on the temperature of particle 
production $T_{\rm f}$. The most important parameter in such 
a theoretical  evaluation is the initial temperature at which the deconfined
phase is created. As noted above,  we estimated this temperature 
at $T_{\rm in}=320$ MeV \cite{analyze2}.
Further uncertainty of the calculation arises from the strange quark mass 
taken  here  to be $m_{\rm s}(1\mbox{\,GeV})=200$ MeV. We recall that 
the strength of the 
production rate is now sufficiently constrained by the measurement of 
$\alpha_{\rm s}(M_Z)$.
We choose  a geometric size which comprises a baryon number $B= 300$
 at $\lambda_{\rm q}\simeq 1.5$, and have verified that
 our result will be little dependent on small
variations in $B$. We show in Fig.\,\ref{gammTf} how the computed
$\gamma_{\rm s}$ depends
on formation temperature $T_{\rm f}$. The cross to the right
shows our  fitted value from line 1 or 2 in table~\ref{t1}. 
It is gratifying to see that it is consistent with the 
theoretical expectation  for early formation of the strange 
(anti)baryons. The relative smallness of $\gamma_{\rm s}$,
despite the hight strangeness yield, is 
clearly related to the high temperature of particle production.

The relatively high value of temperature $T_{\rm f}$ we obtained in the 
 QGP reaction picture with primordial particle emission is in qualitative 
agreement with the experimental inverse slope data we discussed in section 2.
We are thus led to the conclusion that 
 as far as the fitted temperatures and slopes are 
concerned, it is possible that  the high $m_\bot$ 
strange (anti)baryons we have described could have been emitted directly 
from a primordial (deconfined) phase before it evolves into final 
state hadrons.

Is QGP primordial emission hypothesis also consistent with the chemical fugacities 
we have obtained? The chemical condition is 
fixed to about 5\% precision, and there is 40\% anti-correlation
between the two fugacities $\lambda_{\rm q}$ and $\lambda_{\rm s}$. The
information that
$\lambda_{\rm s}\ne1$ is contained in at least two particle
abundances; arbitrary manipulation of the reported yields of one particle 
abundance did not reduce the value $\lambda_{\rm s}$ to unity. 
Since $\lambda_{\rm s}\ne 1$ 
by  4\,s.d.  it is highly unlikely that $\lambda_{\rm s}=1$ 
is found after more data is studied. While one naively expects 
$\lambda_{\rm s}^{\rm QGP}=1$, to assure the strangeness balance 
$\langle s-\bar s\rangle=0$, there must be 
a small deviations from this value, even if the emitted particles were to
reach asymptotic distances without any further interactions: 
in presence of baryon density the deconfined state is not
fully symmetric under interchange of particles with antiparticles. 
A possible mechanism to distinguish the strange and anti-strange quarks 
arises akin to the effect considered for the $K^-/K^+$ asymmetry  in baryonic 
matter \cite{Shu92,Go92}: there is asymmetric scattering  strength on 
baryon density $\nu_{\rm b}$
 which causes presence of a mean effective vector potential
$W$. Similarly, strange quark interaction with baryon density would lead to a
dispersion relation 
\begin{equation}\label{Ws}
E_{\rm s/\bar s}=\sqrt{m_{\rm s}^2+p^2}\pm W \, ,
\end{equation}
and this requires in the statistical approach that the 
Fermi distribution for strange and antistrange 
quarks acquires a  compensating fugacity 
$\lambda_{{\rm s},\bar {\rm s}}=e^{\pm W/T}$ 
to assure strangeness balance in the deconfined phase. In linear
response approach $W\propto\nu_{\rm b}$  consistent with both $W$  and
baryon density $\nu_{\rm b}=(n_{\rm q}-n_{\bar {\rm q}})/3$, 
being fourth component of a Lorenz-vector. 
It is clear for intuitive reasons, as well as given experimental 
observations, that the baryon stopping and thus density increases
considerably comparing the S and Pb induced reactions in the energy domain
here considered. We also recall that in S--W reactions 
 $\lambda_{\rm s}^{\rm S}\simeq 1.03\pm 0.05$ \cite{analyze}. 
Should in the dense matter fireball the 
baryon density $\nu_{\rm b}$ grow by factor 2--4 
as the projectile changes from S to Pb, this alone would
consistently explain  the appearance of the value $\lambda_{\rm s}=1.14\pm0.04$
obtain using $W\propto\nu_{\rm b}$ scaling. 
It is worth noting that the value $W\simeq 38$\,MeV suffices here. Note also
that the Coulomb potential effect on the charge of the strange quarks is
of opposite magnitude and about 1/5--1/6 of the here required strength.

\subsection{Late emission scenario: HG with or without QGP?}
A generally favored picture of particle production involves flow expansion 
of the primordial phase till a transition temperature is reached, 
at which time the final state hadrons are produced, and 
soon thereafter freeze out. These particles may
directly reach a detector or reequilibrate and appear to the
observer as if emitted from a HG phase, except that 
entropy/strangeness  excess effect should remain. 
In order to force our data fit to converge to such a late particle 
production scenario we introduce the experimental result,
which was essential for such an argument in the S--Pb induced 
reactions. The quantity of interest is ratio of particle yields 
between particles of very different mass. Thus
in lines 3, 4 in table~\ref{t1} we include in the fit also our
estimate of the hyperon to kaon ratio Eq.\,(\ref{klratio}), thus 
altogether we now fit 9 data points.

Our approach in the third line corresponds to a freeze-out from
`cold' QGP phase, in that we allow the 
abundance parameters $\gamma_{\rm s}$ and $R_{\rm C}^{\rm s}$ 
to deviate from HG equilibrium values. 
We note that this cold-QGP alternative has a 
very comparable statistical significance as the hot-QGP.
Given the low temperature and high  $m_\bot$ 
inverse slopes we must have considerable transverse flow.
The  computed flow velocity at freeze-out is $v_{\rm f}=0.51$c. 
This is just below the  relativistic sound velocity 
$v_{\rm s}=1/\sqrt{3}=0.58$\,, which we have assumed.
 In  Fig.\,\ref{gammTf} the cross to the left shows the 
result of the fit we just described; allowing for potentially smaller expansion 
velocity and all the above discussed uncertainties in the 
computation, this result must also be seen as a very good agreement 
between the result of data fit and the theoretical calculation.
This also means that we cannot distinguish in the present data between 
early formation of strange antibaryons and an expansion model followed
by direct global hadronization. 

Since the fitted values of $\gamma_{\rm s}$ and $R_{\rm C}^{\rm s}$ allow the HG 
equilibrium, we attempt such a fit in line 4, where the particle
yields are fitted constrained for HG equilibrium, and
we use strangeness conservation to evaluate the strangeness fugacity 
$\lambda_{\rm s}$. We show the result of the fit in the last line 
of  table~\ref{t1} and in particular we note:
\begin{eqnarray}
T_{\rm f}&=155\pm 7\,\mbox{MeV},\ &\rightarrow v_\bot
\simeq 0.5\simeq v_{\rm s};\nonumber\\
\lambda_{\rm q}&=1.56\pm0.09,\hspace*{0.25cm} \ &\rightarrow\  \lambda_{\rm s}=1.14;\nonumber\\
\chi^2/9&=0.84,\hspace*{1.4cm} \ &\rightarrow \ \ \mbox{C.L.}>60\%\,.
\end{eqnarray}
We recall that the baryochemical potential is given in 
terms of $T$ and $\lambda_{\rm q}$, specifically $\mu_{\rm b}=3T\ln \lambda_{\rm q}$, and we 
find $\mu_{\rm b}=204\pm10$ MeV in this hadronic gas condition. 
We note that while the quality of the fit has degraded, it 
still has considerable statistical significance. 
It would appear to be a `good' fit to the naked, unequiped eye. 

\begin{figure}[tb]
\vspace*{-1.6cm}
\centerline{\hspace*{-.8cm}
\psfig{width=8cm,figure=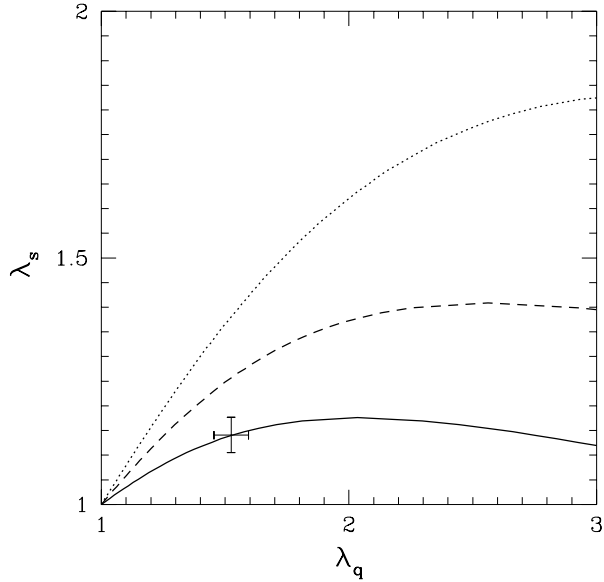}
}
\vspace*{-1.4cm}
\caption{Strangeness conservation constraint in HG
as function of freeze-out $\lambda_{\rm q}$: the 
lines correspond to different freeze-out temperatures 
$T_{\rm f}$ (solid 160 MeV, dashed 140 MeV and dotted 120 MeV).
The cross corresponds to
the chemical freeze-out we determined above.
\label{s-sbarHG}
}

\end{figure}

Unlike S-induced reactions, the  hypothesis of an equilibrated HG 
in the final state cannot be easily 
discarded in Pb--Pb collisions since the 
thermo-chemical parameters are extraordinarily consistent 
with  this hypothesis and the principle of strangeness conservation, as we 
illustrate in
Fig.\,\ref{s-sbarHG}. Here, the cross corresponds to the fitted properties 
of the particle source, while the lines correspond to the constraint of the
HG gas source to yield $\langle s-\bar s\rangle=0$ at finite baryon 
density represented by the value of $\lambda_{\rm q}$. Thus the strange quark 
fugacity is in general not zero and the cross falls just 
on freeze-out at $T=160$ MeV when the meson-baryon abundance is equilibrated.
The slight difference in freeze-out value to results shown in last line of 
table~\ref{t1} is result of using the WA97 $\bar\Lambda/\Lambda$ value Eq.\,(\ref{rat1}) rather than the averred value Eq.\,(\ref{rat2}).

If HG is indeed present in the final state, 
 the proper interpretation of these data, and the likely reaction
scenario, compatible with our earlier work on S induced reactions 
\cite{acta96} is as follows; the relatively large fireball of 
dense and deconfined matter disintegrates and produces 
dense, confined hadronic gas in which strange particles have time to 
rescatter and to establish relative chemical equilibrium. 
A possible test of this hypothesis would be to see variation of 
the chemical parameters  as the size of the fireball changes 
with impact parameter (centrality of collision) 
since reequilibration should diminish for small reaction volume.
However, such data are presently not available, and 
there is no indication that indeed a change of the strange (anti)baryon 
yields occurs as the centrality of the interaction is reduced. 
On the other hand, the specific entropy and strangeness should
comprise a signal of some new physics should
formation and expansion of QGP phase, followed by 
reequilibration into HG phase, and freeze-out have occurred.
We will now consider the magnitude of these effects:\\

\noindent{\bf Strangeness reequilibration}\\
When  HG emerges from initial dense  QGP phase, the
number of strange quark pairs does not change, but the 
phase space density of strangeness changes, since the phases
are different. Because the HG phase has generally a smaller
phase space density of strangeness than QGP, to conserve 
strangeness, there will be a jump in the phase space occupancy 
$\gamma_{\rm s}$ during the transformation of QGP into HG, as there 
is a jump in the strange quark fugacity. The important point is 
that this could lead to significantly 
overpopulated  HG phase ($\gamma_{\rm s}>1$). This phenomenon 
can be easily quantified as follows:
the observed value of $\gamma_{\rm s}^{\rm HG}$ is related to the pre-phase
change value  $\gamma_{\rm s}^{\rm QGP}$ by introducing the enhancement 
factor we wish to determine:
\begin{equation} \label{F1}
\gamma_{\rm s}^{\rm HG}\equiv F_\gamma \gamma_{\rm s}^{\rm QGP}
\end{equation}
A simple way to compute the value of the saturation enhancement 
factor $F_\gamma$ is to study the
abundance of strangeness per baryon number 
before and after phase transition.
\begin{equation}\label{F2}
F_\gamma =\frac{s/b|_{QGP}}{s/b|_{HG}} 
    = \frac{\gamma_{\rm s}^{\rm QGP}}{\gamma_{\rm s}^{\rm HG}} 
                    f(T_{\rm f},\lambda_{\rm q},\gamma_{\rm s}^{\rm HG})\,.
\end{equation}
The last expression arises as follows: 
on the QGP side the abundance of 
strangeness is to a good approximation proportional 
to $\gamma_{\rm s}^{\rm QGP}$ and is the 
integral of the strange quark phase 
space, we evaluate it assuming that 
$m_{\rm s}/T_{\rm QGP}\simeq 1$. There is no
dependence on chemical properties of 
the plasma. On HG side, at freeze-out 
we have to evaluate the strangeness abundance 
from the strange particle 
partition function given in Eq.\,(16) of \cite{analyze1}, 
supplemented by the
now relevant term comprising
 $s\bar s$--$\eta,\ \eta',\ \phi$ states
and their resonances. The sum includes 
a terms proportional to 
$(\gamma_{\rm s}^{\rm HG})^n$, with n=1, 2, 3, 
indicating strangeness content of hadrons. 
The leading kaon and hyperon term is proportional 
to $\gamma_{\rm s}^{\rm HG}$ and hence
we have above result, Eq.\,(\ref{F2}). 
We thus obtain, combining Eqs.\,(\ref{F1}) and Eq.\,(\ref{F2}),
\begin{equation}\label{F3}
F_\gamma^2=F_\gamma
    \frac{\gamma_{\rm s}^{\rm HG\ }}{\gamma_{\rm s}^{\rm QGP}}=
\left.\frac{s}{\gamma_{\rm s}b}\right|_{QGP}\cdot \left.\frac{\gamma_{\rm s} b}{s}\right|_{HG}\,, 
\end{equation}
where the right hand side now compares the properties of the two 
phases at the boundary between them and we can evaluate it using 
the theoretical equations of state. In analyzes of an 
experiment we would take the  freeze-out parameters 
determined by the fit to data.

\begin{figure}[tb]
\vspace*{-1.6cm}
\centerline{\hspace*{-.8cm}
\psfig{width=8cm,figure=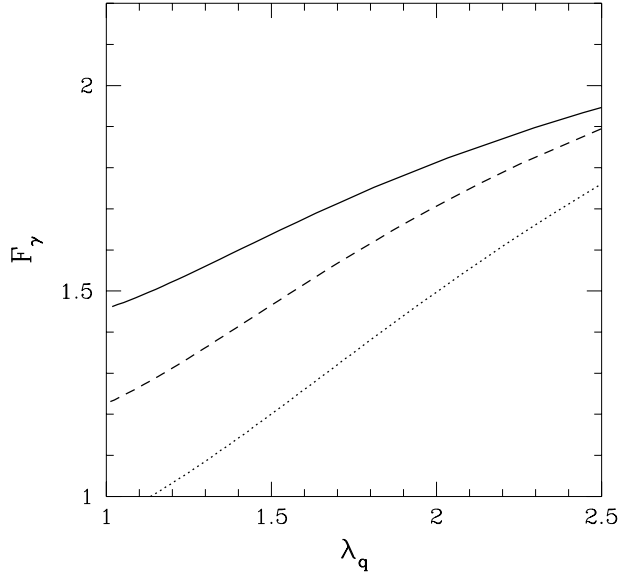}
}
\vspace*{-1.4cm}
\caption{Strangeness phase space enhancement factor as function of
HG freeze out $\lambda_{\rm q}$ for $T_{\rm f}=160$ (solid line),
$T_{\rm f}=140$ (dashed line), $T_{\rm f}=120$ (dotted line). Computed
for $\gamma_{\rm s}^{\rm HG}=1$, $m_{\rm s}/T_{\rm QGP}=1$, including in HG phase
kaons, hyperons, cascades, $\eta$, $\phi,\, \Omega$, and imposing 
strangeness conservation constraint to determine~$\lambda_{\rm s}$. 
\label{fgamma}
}

\end{figure}
We show, 
in Fig.\,\ref{fgamma},  the strangeness enhancement factor as function of 
$\lambda_{\rm q}$ for several freeze-out temperatures $T_{\rm f}=160,\,140,\,120$ MeV,
with $\lambda_{\rm s}$ fixed by strangeness conservation constraint. We see 
that $F_\gamma$ varies  typically between 1.5 and 2, and is specifically 
1.6 for the parameter rage  of Pb--Pb collisions here discussed. 
This means that observing the value
$\gamma_{\rm s}\simeq 1$ really means an underlying value $\gamma_{\rm s}^{\rm QGP}\simeq 0.6$.
Conversely, should we be able to create a  longer lived or hotter QGP state 
we could expect to observe in the HG phase $\gamma_{\rm s}^{\rm HG}$ as large as
1.5--2. Such over-saturation of the phase space would 
be a rather strong smoking gun
pointing to the formation of the QGP phase.\\ 

\noindent{\bf Entropy and particle excess}\\
Another way to argue for the formation of QGP  in early stages of
an expansion  scenario of the fireball is to measure 
 the specific entropy experimentally,
for example by measuring the quantity
\begin{equation}    
   D_{\rm Q} \equiv \left({\displaystyle{dN^+\over dy} - {dN^- \over dy}}
\right) \left/\left( {\displaystyle{dN^+\over dy} + {dN^- \over dy}}\right) \,,
\right.\end{equation}
which we have shown to be a good measure of the entropy content \cite{analyze}.
We note that in the numerator of $D_{\rm Q} $ the charge of particle
pairs produced cancels and hence this value is effectively a measure of
the baryon number, but there is a significant correction arising from
the presence of strange particles. The denominator is a measure of the
total multiplicity --- its value is different before or after disintegration
of the produced unstable hadronic resonances. Using as input the
distribution of final state particles as generated within the hadron gas
final state it is found \cite{entropy} that $D_Q\cdot S/B$  is nearly
independent of the thermal  parameters and varies between 4.8, before
disintegration of the  resonances,  to 3 after disintegration. 

\begin{figure}[tb]
\vspace*{-1.6cm}
\centerline{\hspace*{-.8cm}
\psfig{width=8cm,figure=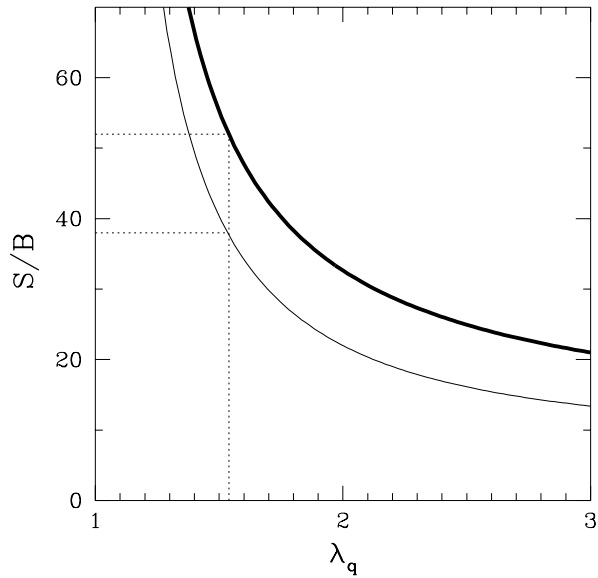}
}
\vspace*{-1.4cm}
\caption{\small
QGP (thick line) and HG (thin line, $T=155$ MeV) entropy per baryon 
$S/B$ as function of light quark fugacity $\lambda_{\rm q}$. Dotted lines
guide the eye for the here interesting values.
\label{entro1}
}
\end{figure}

To obtain a measure of the particle
excess, we show in  Fig.\,\ref{entro1} the specific entropy  per baryon
$S/B$ content in  dense hadronic matter  as a function of  light 
quark fugacity $\lambda_{\rm q}$. The thick line addresses
the deconfined QGP  phase, the thin line the confined HG phase at $T=155$ MeV, 
with the strange quark fugacity $\lambda_{\rm s}$ 
being determined from the strangeness conservation condition. While the 
QGP result is largely independent of temperature (only other,
aside of $T$ dimensioned quantity, is $m_{\rm s}$), the HG result involves 
the values of all hadron masses and hence is
dependent on $T$. The 12 units of entropy 
difference between the two phases for the here interesting range of 
fugacity $\lambda_{\rm q}=1.5$--$1.6$ implies that we should expect an excess of 
about 3 mesons per baryon if the deconfined phase is formed.
We compare with a HG at $T=155$ MeV; 
should the HG phase of interest be hotter, this 
difference between QGP and HG grows, since the baryon density in HG grows 
much faster than entropy for the baryon mass is well below the temperature
range and thus a change in the factor $m_{\rm N}/T$ matters, while the change 
in $m_\pi/T$ is immaterial. In other work HG phase at 190 MeV is often 
considered, and there the difference between QGP and HG properties turns out 
to be as large as factor two \cite{entropy}. 

\section{QGP: Year 1997 or 2001?}
In our view, strangeness and entropy results analyzed, here and 
earlier~\cite{acta96}, are most consistent with the hypothesis
that the same type of  matter is formed in both S- and Pb-induced 
reactions, but that quite different `initial' conditions
are reached, specifically  different  energy density and 
 very different baryon density. Moreover, the 
longitudinal flow, clearly visible in the S--S 200A GeV, in
{\it e.g.}, $\Lambda$-rapidity spectra, is not seen
in Pb--Pb reactions (see Fig.\,\ref{NA49LAL}), but there could be 
strong transverse/radial flow driven by high internal pressure 
in the fireball, 
required to connect  the low freeze-out temperature 
(see third and fourth  fit of table  \ref{t1})  to 
the high inverse $m_\bot$ slope, see Eq.\,(\ref{shiftT}). 
We also expect that the much greater
volume occupied by the initially deconfined phase  
should enhance the reequilibration
process of hadrons after freeze-out, furthering the 
approach to chemical equilibrium particle
yields.

Given these considerable differences in reaction dynamics,
it is surprising that  the strange antibaryon yields from Pb--Pb 
reactions appear to be, at least on first sight, similar to the S--W/Pb or even 
S--S results. Our understanding and  explanation
of this phenomenon is that the abundance (but not necessarily the spectrum) 
of relatively high $m_\bot$ 
strange (anti)baryons is originating in a similar 
stage of a hot matter fireball. To this point we note that
several {\it ad-hoc} mechanisms 
(such as, {\it e.g.}, `ropes', `string fusion', {\it etc.})
were proposed in literature aiming
to explain the strange antibaryon yields in S--S/Ag/W/Pb reactions, and
which should in case of Pb--Pb reactions yield even a greater strange 
antibaryon production anomaly, not seen in the 
experimental results known to us and described here. 

At this point we wish to address
another important experimental result which indeed
shows considerable differences between S--A and Pb--Pb
data \cite{these}: the rate of absorption of the $J\!/\!\Psi$, 
per unit of  size of the interaction region,
 is systematically enhanced when the reaction region is sufficiently 
large, as measured by transverse energy. 
We wish to interpret that exciting result in a slightly 
different manner  than has been proposed before \cite{Gon96},
in order to reconcile it with our claim that S-- and Pb-- induced 
reaction lead to the same type of (deconfined) dense primordial 
hadronic matter.

Seen in the CM-frame of reference for the Pb--Pb system, 
the produced $J\!/\!\Psi$ is almost `at rest' compared to the 
rapidly evolving, `light' hadronic matter. In an idealized picture, 
$J\!/\!\Psi$ sits still as it is banged on initially by quarks and gluons,
and if reequilibration is occuring, also by individual hadrons. One 
of the results of this reaction picture is that we expect the 
$m_\bot$ spectrum of $J\!/\!\Psi$ to become thermal, 
since the particles hitting it 
are thermal; the other is that the observed absorption
of $J\!/\!\Psi$ is not a function of the size of the fireball, as 
 $J\!/\!\Psi$ does not take a straight path 
out of the production point through the surrounding matter. If it is 
moving out, it would be certainly
not straight given the interactions with
the medium; despite its great mass compared to massless 
quarks and gluons, the nonrelativistic thermal momentum of  
$J\!/\!\Psi$: $p=\sqrt{3mT}$ is, at $T=250$ MeV, just twice as large as the 
thermal relativistic momentum of the light quarks and gluons, 
$p\simeq3T$, which assures rapid thermalization  of $J\!/\!\Psi$.
In this approach $J\!/\!\Psi$ absorption depends on
how long it is exposed to interactions with 
hadronic matter in its various forms, perhaps most importantly
with the color screening QGP phase, and 
 the `anomalous' suppression of $J\!/\!\Psi$
in central Pb--Pb reactions simply implies that the large, 
dense hadronic fireball 
has an extended lifespan compared to peripheral Pb--Pb reactions or
S- and p-induced reactions. Critical phenomena such as 
development of a `frozen' HG envelope (expanding considerably slower) 
around  the deconfined region could lead to the reported
sudden change in $J\!/\!\Psi$ suppression with $E_\bot$.

We would expect an equally sudden change as
function of $E_\bot$ of the chemical parameter $\lambda_s$ 
since formation of a `frozen' HG surface should also lead to 
 strange (anti)baryon  reequilibration in this 
 matter envelope. 
A way to verify this hypothesis would be to show that,
as function of Pb--Pb collision centrality, the strange quark fugacity
drops (quite suddenly) from the HG equilibrium value to the 
QGP direct emission value $\lambda_{\rm s}\to 1$, just at the same
critical $E_\bot$ noted in study of $J\!/\!\Psi$ suppression.

We conclude that the appropriate source of strange particles
in Pb--Pb collisions is an expanding QGP fireball, 
that undergoes a reequilibrating phase transformation,
capable to delay its dissociation. This 
scenario represented in table \ref{t1}, line 3, is as argued above
qualitatively consistent
with the anomalous $J\!/\!\Psi$ suppression, provided that we do not
model $J\!/\!\Psi$ as escaping the reaction region, but instead  see
matter flowing out of the central formation region of the $J\!/\!\Psi$. 

Let us stress again that in our here presented work, 
we did assume that thermal quark-gluon degrees
of freedom are at origin of many of the hadronic particle production
phenomena in relativistic hadron reactions. 
Many simple, but subtle experimental observations point in this 
natural direction. For example, all so far studied $m_\bot$ spectra 
in S- and Pb-induced reactions have the same shape for strange
baryons and antibaryons of the same kind, and even for different 
kinds, where comparison can be made
in same range of $m_\bot$. This is not an accident, but 
result of either complete thermal equilibrium, or of
their origin in a thermal source
composed of their constituents (quarks). Only a 
thermal quark liquid can deliver this result naturally. 

Similarly, we take the presence of near
chemical equilibrium of strangeness to be a signal of primordial QGP 
phase. The phase space occupancy factor 
expected in the QGP phase is of magnitude 0.6 and is enhanced by the lower 
strangeness density in the HG phase by a factor 1.5 to reach unity. 
The chemical parameters that are observed lead to abundance anomalies such as
$\bar\Lambda /{\bar p}>1$. There is a priori no reason for a
HG state to reach condition amenable to this result, should it not 
arrive from a QGP state.

All told, we believe that the most simple an consistent 
reaction picture involves formation of deconfined phase 
of  hadronic matter both in 
S- and Pb-induced reactions. The difference between both 
cases is that the former leads to a small enough fireball
that can rapidly
disintegrate under influence of the longitudinal flow and 
without forming an intermediate fully equilibrated 
HG phase. The Pb--Pb reactions, comprising five time 
the amount of matter, considerably higher energy density 
and much less longitudinal flow,
appear to undergo a more protracted  evolution history,
and in particular they appear to pass through a stage of 
reequilibrated HG, before the strange antibaryons 
decouple (freeze-out) and stream freely to the detectors.

We have no idea how to describe the relativistic heavy ion hadron
spectra data without invoking thermal quark-gluon fluid, commonly 
referred to as QGP. However, we have many more experiments and
theoretical analysis to perform to be sure that we understand the 
reaction history and the properties of the bizarre liquid
of quarks and gluons. The answer thus is, QGP is arriving in 1997,
but needs to be properly dressed for 
more general presentation to the public.\\

\noindent\section*{\bf Acknowledgments:} One of us (J.R.) would like to thank the  organizers
of the XXXVII Zakopane Summer School, and in particular Jan Czy\.zewski and 
Jacek Wosiek, and Marja Czy\.z, for their great
Polish  hospitality during a most fruitful and interesting meeting.



\end{document}